\documentclass[12pt]{article}

\usepackage[margin=1in,headsep=.3in]{geometry}
\usepackage{fancyhdr}
\usepackage{amsmath, amssymb, bm}
\usepackage{graphicx}
\usepackage{caption}
\usepackage{xcolor}
\usepackage{hyperref}
\usepackage{multicol}
\usepackage{setspace}
\usepackage{indentfirst}
\usepackage[title]{appendix}
\usepackage{dirtytalk}
\usepackage[hang,flushmargin]{footmisc}
\usepackage{biblatex}
\usepackage[font=footnotesize]{caption}
\title{\vspace{3cm} 
Gravitational Waves from Simulated Mergers \\[0.5em]
of 2 and 3 Black Holes }

\author{
\vspace{0.5cm}
Faisal Alotaibi \\[0.8em]
\textnormal{Master of Science Project}\\[0.4em]
\textnormal{Department of Mathematical and Statistical Sciences}\\[0.4em]
\textnormal{University of Colorado Denver}\\[1.2em]
\small Advisor: Dr. Aim\'e Fournier
}
\date{July 30, 2025}

\begin{document}
\maketitle
\thispagestyle{empty}
\newpage
\setcounter{page}{1}
\section*{Abstract}
In this project, we simulate the collision of two and three black holes using NRPy+ (`Python-based code generation for numerical relativity and beyond') module and BSSN (Baumgarte-Shapiro-Shibata-Nakamura) formulation, and extract the resulting gravitational waveforms.  Using Brill-Lindquist initial data and sixth-order finite differences, we evolve the system using the BSSN formulation and compute the gravitational-wave signal via the Weyl scalar $\psi_4$. To assess numerical error, we plot the Hamiltonian constraint and observe that constraint violations are significantly higher in the three-black-hole collision. Unexpected gravitational recoil is also detected, which may influence waveform extraction and that is left for further investigation. Despite the limitations in computational resources imposed by the Google Colab, we successfully model the merger of a binary black hole system, and we were able to extract the corresponding gravitational waves.

\vspace{3mm}
\section*{Introduction}
In 1905, Albert Einstein introduced the theory of Special Relativity to better understand the connection between space and time. Ten years later, he extended this idea to include gravity, resulting in the theory of General Relativity (GR), which describes how mass and energy shape the structure of space and time itself \cite{carroll}. While we often think of space as three-dimensional and time as separate, GR combines them into a single framework called spacetime. Unlike Newton’s view of gravity as a force between objects, GR explains gravity as the curvature of spacetime caused by the presence of mass and energy.

\vspace{3mm}
Einstein's field equations (EFE) in GR relate the curvature of spacetime to the distribution of matter and energy. The EFE are ten nonlinear partial differential equations, making it difficult to solve them analytically. Thus, solving them numerically comes as an alternative way to deal with the complexity of the solutions. Therefore, the field numerical relativity (NR) has become widely used over the years. 

\vspace{3mm}
In NR, there are some challenges when using numerical methods such as maintaining numerical stability over long simulations. ADM (Arnowitt-Deser-Misner) formalism was one of the first formulations  of EFE, but the solutions become unstable in long term simulation \cite{Palenzuela}. BSSN formalism is a modified version of ADM formalism, and it is more stable on the long run simulation since it is strongly hyperbolic and well-posed, unlike the weakly hyperbolic ADM system \cite{Baumgarte}. Thus, maintaining stability is crucial to achieve a reliable simulation for our solutions.

\vspace{3mm}
There are some Python-based modules and software tools commonly used to study and simulate phenomena in numerical relativity, including black holes, gravitational waves, and neutron stars. Supercomputers are mostly used due to the high computational demands of solving EFE. Unlike other software modules, NRPy+ requires no expensive Mathematica or Maple license to function. Instead, it relies on SymPy, the standard Python library for symbolic algebra, which is used within NRPy+ to generate functions and definitions that support calculations in the main simulation files, thereby lowering the barrier to entry for new users \cite{etienneBHaH}.

\vspace{3mm}
In this work, we take advantage of NRPy+ accessibility to move beyond typical binary systems. While most numerical relativity studies focus on single or binary event systems, in this project we explore more complex dynamics of three colliding black holes. Using NRPy+, we simulate the collision and try to extract the resulting gravitational waveforms via the Weyl scalar $\psi_4$. We then compare our results with a binary system and discuss the challenges we encountered when implementing the three black hole simulation.

\vspace{3mm}
Throughout this paper, Latin indices $a, b, c, \ldots$ denote 4-dimensional spacetime components, with the zeroth component representing time. While the Latin indices $i, j, k, \ldots$ are used for 3-dimensional spatial components. We adopt geometrized units, setting the gravitational constant and the speed of light to unity: $G = 1$ and $c = 1$, so that all quantities have dimensions of length. In addition, We use the Einstein summation convention, which implies summation over repeated indices. 

\vspace{3mm}
\subsubsection*{Derivation of BSSN Formalism and the 3+1 Decompositions}
As we mentioned above, BSSN formalism is based on the ADM formalism, which splits the 4-dimensional spacetime manifold into separate spatial and temporal components, $3+1$ spacetime, reformulating the Einstein equations into a set of evolution equations for 3-dimensional geometric fields. The Einstein field equations are
\begin{align} \label{eq:EFEs}
    G_{ab} \equiv R_{ab} - \frac{1}{2} R g_{ab} = 8\pi T_{ab},
\end{align}
where $G_{ab}$ is the Einstein tensor, $R_{ab}$ is the Ricci tensor, $R$ the Ricci scalar, $g_{ab}$ the spacetime metric, and $T_{ab}$ the stress-energy tensor, see Appendix \ref{appendix:A}.  Note that in this project we are going to work on vacuum black hole spacetimes, i.e. $T_{ab}=0 $, except at the puncture\footnote{The puncture represents the point at the the black hole's center where the conformal factor becomes singular.} \cite{Palenzuela}. The main idea behind BSSN is to use conformal transformations and introduce extra variables, auxiliary variables, to reduce numerical instabilities that occur when solving the original ADM equations.
\begin{figure}[ht]
    \centering
    \captionsetup{width=0.8\linewidth}
    \includegraphics[width=80mm]{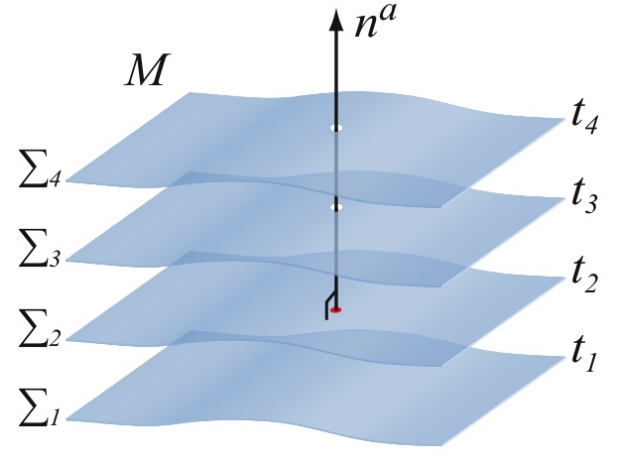} 
    \caption{\footnotesize Foliation of spacetime into spatial hypersurfaces $\Sigma_t$ labeled by coordinate time $t$. The unit normal vector $n^a$ is orthogonal to each slice. [Adapted from \cite{Baumgarte}]}
    \label{fig:3+1}
\end{figure} 

\vspace{3mm}
In the ADM formulation we have 4 main variables. The 3-spatial metric $\gamma_{ij}$ describes the intrinsic geometry of the spatial slices. $K_{ij}$, eqation (\ref{eq:extrinsic}), is the extrinsic curvature, which characterizes how a spatial hypersurface $\Sigma_t$ is embedded in the full 4-dimensional spacetime $\textbf{M}$. The lapse function $\alpha$ relates coordinate time $t$ to the proper time\footnote{The time as measured by a clock moving along with an observer.} $\tau$ of a normal observer via $d\tau = \alpha \, dt$. Finally, the shift vector $\beta^i$ that describes how spatial coordinates are shifted between slices, see Appendix \ref{appendix:A} for more details.

\vspace{3mm}
The ADM formalism has two different groups of equations. First group is a set of first-order-in-time partial differential equations for $\gamma_{ij}$ and $K_{ij}$, which we call the evolution equations. These are derived directly from the Einstein field equations. The second group is called the constraint equations, the Hamiltonian and the momentum constraints. These equations must be satisfied by the initial data on a spatial slice. If satisfied initially, the evolution equations should preserve them. Evolving a system using ADM evolution equations would result in numerical instabilities in strong field regions. 

\vspace{3mm} 
The BSSN formulations address these issues by conformally transferring the ADM variables. $\gamma_{ij}$ is decomposed into a conformal metric $\bar{\gamma}_{ij}$ and the conformal factor $e^{2\phi}$. Thus, 
\begin{align}
    \gamma_{ij} = e^{4\phi} \bar{\gamma}_{ij}.
\end{align}

\vspace{3mm}
We split $K_{ij}$ into two parts, its trace 
\begin{align}
    K = \gamma^{ij} K_{ij},
\end{align}
and the conformal trace-free 
\begin{align}
    \bar{A}_{ij} = e^{-4\phi} \left(K_{ij} - \frac{1}{3}\gamma_{ij}K\right).
\end{align}

The spatial Christoffel symbols $\Gamma_{jk}^i$, equation (\ref{eq:Christoffel}) are derived from the derivatives of $\gamma_{ij}$ which are involving second derivatives that can be numerically noisy. We introduce a new variable, the conformal connection function $\bar{\Lambda}^i$, which solve this problem, 
\begin{align}
    \bar{\Lambda}^i \equiv \bar{\gamma}^{jk} \bar{\Gamma}^i_{jk},
\end{align}
where $\bar{\Gamma}^i_{jk}$ are the Christoffel symbols computed from the conformal metric $\bar{\gamma}_{ij}$. This avoids the direct calculation of second derivatives of
the metric. Then we derive the BSSN evolution equations by substituting the above conformal decompositions into the original ADM evolution equations and get
\newpage

\begin{subequations} \label{eq:evolution}
\begin{align}
  \partial_{\perp} \bar{\gamma}_{i j} {} = {} & \frac{2}{3} \bar{\gamma}_{i j} \left (\alpha \bar{A}_{k}^{k} - \bar{D}_{k} \beta^{k}\right ) - 2 \alpha \bar{A}_{i j} \; , \\
  \partial_{\perp} \bar{A}_{i j} {} = {} & -\frac{2}{3} \bar{A}_{i j} \bar{D}_{k} \beta^{k} - 2 \alpha \bar{A}_{i k} {\bar{A}^{k}}_{j} + \alpha \bar{A}_{i j} K \nonumber 
   + e^{-4 \phi} \left \{-2 \alpha \bar{D}_{i} \bar{D}_{j} \phi + 4 \alpha \bar{D}_{i} \phi \bar{D}_{j} \phi \right .  \\
    & \left . + 4 \bar{D}_{(i} \alpha \bar{D}_{j)} \phi - \bar{D}_{i} \bar{D}_{j} \alpha + \alpha \bar{R}_{i j} \right \}^{\text{TF}} \; ,  \\
  \partial_{\perp} \phi {} = {} & \frac{1}{6} \left (\bar{D}_{k} \beta^{k} - \alpha K \right ) \; , \\
  \partial_{\perp} K {} = {} & \frac{1}{3} \alpha K^{2} + \alpha \bar{A}_{i j} \bar{A}^{i j}  
   - e^{-4 \phi} \left (\bar{D}_{i} \bar{D}^{i} \alpha + 2 \bar{D}^{i} \alpha \bar{D}_{i} \phi \right ) \; , \\
  \partial_{\perp} \bar{\Lambda}^{i} {} = {} & \bar{\gamma}^{j k} \hat{D}_{j} \hat{D}_{k} \beta^{i} + \frac{2}{3} \Delta^{i} \bar{D}_{j} \beta^{j} + \frac{1}{3} \bar{D}^{i} \bar{D}_{j} \beta^{j} \nonumber 
   - 2 \bar{A}^{i j} \left (\partial_{j} \alpha - 6 \partial_{j} \phi \right ) + 2 \bar{A}^{j k} \Delta_{j k}^{i}  \\
  & -\frac{4}{3} \alpha \bar{\gamma}^{i j} \partial_{j} K \; ,
\end{align}
\end{subequations}
where the \text{TF} superscript denotes the trace-free part \cite{BSSNQ}. See Appendix \ref{appendix:A} for details about these equations. The evolution equations \eqref{eq:evolution} are coupled with gauge conditions, the lapse function $\alpha$ and the shift factor $\beta^i$.  The evolution employs the standard moving puncture gauge conditions, the advective $1+\log$ lapse condition which helps avoid black hole singularities, and the advective Gamma-driver shift condition that prevents grid stretching and adjusts the coordinates to adapt to the spacetime dynamics. These gauge conditions are described by the following equations
\begin{subequations} \label{eq:gauge}
\begin{align}
    \partial_0 \alpha &= -2 \alpha K \; ,\\
    \partial_0 \beta^i &= B^{i} \; ,\\
    \partial_0 B^i &= \frac{3}{4} \partial_{0} \bar{\Lambda}^{i} - \eta B^{i} \; ,
\end{align}
\end{subequations}
where $\partial_0$ is an advective time derivative, see Appendix \ref{appendix:A}, $B^i$  is an auxiliary vector, and $\eta$ is a damping parameter that drives $\partial_t \beta^i$ toward zero, causing the shift vector $\beta^i$ to approach a constant value in stationary spacetimes \cite{Schnetter}. In addition to the evolution equations, the BSSN system also includes two constraint equations, the Hamiltonian $\mathcal{H}$ and the momentum $\mathcal{M}^i$ constraints
\begin{subequations} \label{eq:constraint}
\begin{align} 
    \mathcal{H} &= \frac{2}{3} K^2 - \bar{A}_{ij} \bar{A}^{ij}+ e^{-4\phi} \left(\bar{R} - 8 \bar{D}^i \phi \bar{D}_i \phi - 8 \bar{D}^2 \phi\right) = 0 \; ,\\
    \mathcal{M}^i &= e^{-4\phi} \left(\bar{D}_j \bar{A}^{ij} +\bar{A}^{ij}\partial_j \phi -\frac{2}{3} \bar{\gamma}^{ij}\partial_j K\right) = 0 \; .
\end{align}
\end{subequations}
Together, the evolution equations \eqref{eq:evolution}, the gauge conditions~\eqref{eq:gauge}, and the constraints~\eqref{eq:constraint} form what is known as the BSSN formulation of the Einstein equations. 

\vspace{3mm}
At each hypersurface $\Sigma_t$ (see Fig. \ref{fig:3+1}), the constraint equations must be satisfied. Throughout the simulation, these constraints are monitored to evaluate numerical accuracy \cite{IZB}. As will be demonstrated in the results section, violations of the constraints are noticeably larger inside the black hole region. This is expected, as the fields are nonsmooth at the puncture and errors from finite differencing at the singularity become trapped rather than propagate outward \cite{IZB}.

\subsubsection*{Polarizations of the Gravitational Waves}
Consider a small perturbation $h_{ab}$ to a known background solution of Einstein’s equations. While the background can, in principle, be any spacetime, our focus here is on gravitational waves propagating in a spacetime that is approximately Minkowskian, where the background metric approaches the flat metric $\eta_{ab}$. So, we have 
\begin{equation}
g_{ab} = \eta_{ab} + h_{ab}, \qquad |h_{ab}| \ll 1.
\end{equation}

We exploit the freedom in choosing a coordinate system by imposing the Lorenz gauge condition, $\partial^a \bar{h}_{ab} = 0$, where $\bar{h}_{ab}$ denotes the trace-reversed metric perturbation, see equation (\ref{eq:tracerp}). In vacuum, the linearized Einstein field equations reduced to
\begin{equation}  \label{eq:LEFE}
    G_{ab} = -\frac{1}{2} \Box \bar{h}_{ab} = 0 \quad \Rightarrow \quad \Box \bar{h}_{ab} = 0,
\end{equation} 
where $\Box$ is the d'Alembertian operator (\ref{eq:d'A}). Equation (\ref{eq:LEFE}) shows that the linearized Einstein equations take the form of a wave equation for the trace-reversed metric perturbation $\bar{h}_{ab}$. However, $\bar{h}_{ab}$ is not uniquely defined; the residual gauge freedom allows us to impose additional conditions. We now choose the transverse-traceless (TT) gauge \cite{Baumgarte}. In this gauge, $\bar{h}_{ab}$ is purely spatial, so all components involving the time coordinate vanish: $\bar{h}_{a0} = 0$, where 0 denotes the time coordinate, and $x, y, z$ will be the spatial coordinates. Furthermore, the perturbation is traceless, $\bar{h} = \bar{h}^a_{\ a} = 0$, which implies $h = 0$. As a result, the trace-reversed and original perturbations coincide, and we may write $\bar{h}_{ab} = h_{ab}$.

\begin{figure}[ht]
    \centering
    \captionsetup{width=0.8\linewidth}
    \includegraphics[width=150mm]{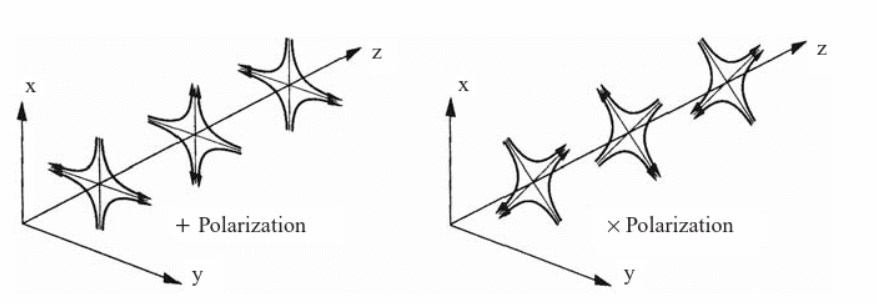} 
    \caption{\footnotesize Lines of force associated with the two polarization states $h_{+}$ and $h_{\times}$ of a linear plane gravitational wave traveling in vacuum in the $z$-direction [Adopted from \cite{Baumgarte}].}
    \label{fig:pluscross}
\end{figure}

\vspace{3mm}
By symmetry, the metric perturbation $h_{ab}$, originally containing 16 components, is reduced to 10 independent components due to its symmetry: $h_{ab} = h_{ba}$. Imposing the Lorenz gauge condition $\partial^a \bar{h}_{ab} = 0$ introduces four constraint equations, reducing the degrees of freedom to six. The transverse-traceless gauge then imposes four additional conditions, three from transversality ($\partial^a h_{a0}^{TT} = 0$) and one from the traceless condition ($h^{TT} = 0$), leaving only two independent components \cite{carroll}. These correspond to the two physical polarizations of the gravitational waves: the “plus” mode $h_+$ and the “cross” mode $h_\times$, given by
\begin{equation} 
h_+ = h_{xx}^{TT} = -h_{yy}^{TT}, \quad 
h_\times = h_{xy}^{TT} = h_{yx}^{TT}. 
\end{equation} 

\vspace{3mm}
\subsubsection*{Construction of $\psi_4$}
In this project,as we mentioned above, we focus on a vacuum black hole spacetimes, where all matter-source terms vanish. As a result, the Ricci tensor becomes ${}^{(4)}R_{ab} = 0$, which implies (see Eq. (1.26) in \cite{Baumgarte}): 
\begin{equation} \label{eq:RC}
    {}^{(4)}R_{abcd} = {}^{(4)}C_{abcd},
\end{equation}
where ${}^{(4)}R_{abcd}$ is the 4-dimensional Riemann curvature tensor, and ${}^{(4)}C_{abcd}$ is the Weyl tensor. Thus, from (\ref{eq:RC}), we see that in vacuum, all information about the curvature of spacetime is contained in the Weyl tensor ${}^{(4)}C_{abcd}$ \cite{Nerozzi}. 

\vspace{3mm}
Note that the 4-dimensional Riemann tensor ${}^{(4)}R_{abcd}$ has 20 independent components. In general, 10 of these reside in the Ricci tensor ${}^{(4)}R_{ab}$, and the remaining 10 are contained in the Weyl tensor ${}^{(4)}C_{abcd}$. These 10 independent components of ${}^{(4)}C_{abcd}$ can be expressed in terms of five complex scalars $\psi_0$, $\psi_1$, $\psi_2$, $\psi_3$, and $\psi_4$ known as the Newman–Penrose (NP) scalars \cite{Baumgarte}. These scalars are constructed by contracting the Weyl tensor with a complex null tetrad.

\vspace{3mm}
The choice of null tetrad affects the values of the Weyl scalars and their physical interpretation. To isolate the gravitational wave content, we identify a special class of tetrads called the transverse frame, in which $\psi_1 = \psi_3 = 0$. Within this class, the quasi-Kinnersley frames, the Weyl scalars $\psi_0$ and $\psi_4$ can be interpreted as representing the ingoing and outgoing gravitational radiation, respectively, while $\psi_2$ encodes the longitudinal part of the gravitational field associated with the mass and angular momentum of the spacetime \cite{Nerozzi}. We construct a null tetrad by choosing two real vectors $l_a$ and $k_a$, and complex conjugate null vectors $m_a$ and $\bar{m}_a$. These vectors satisfy the normalization conditions $l^a k_a = -1$ and $m^a \bar{m}_a = 1$, with all other inner products vanishing.

\vspace{3mm}
We focus on the Weyl scalar $\psi_4$, as it encodes the outgoing gravitational radiation. It is defined as
\begin{equation} \label{eq:psi4C}
    \psi_4 = - {}^{(4)}C_{abcd} \, k^a \bar{m}^b k^c \bar{m}^d.
\end{equation}

\vspace{3mm}
For simplicity, we adopt a tetrad constructed from an orthonormal basis aligned with spherical polar coordinates, as defined in Eq. (9.122) \cite{Baumgarte}. Moreover, by applying Eq. (\ref{eq:RC}), the expression for $\psi_4$ in Eq. (\ref{eq:psi4C}) can be equivalently written in terms of the Riemann tensor; see Eq. (9.123) in \cite{Baumgarte}.

\vspace{3mm}
Since we are focusing on gravitational waves propagating in a nearly flat spacetime, the linearized Riemann tensor is given by
\begin{equation} \label{eq:linearized Riemann}
{}^{(4)}R_{abcd} = \frac{1}{2} \left( 
\partial_a \partial_d h_{bc} 
+ \partial_b \partial_c h_{ad} 
- \partial_b \partial_d h_{ac} 
- \partial_a \partial_c h_{bd} 
\right).
\end{equation}

Here, we work in the transverse-traceless gauge, a radially propagating gravitational wave has only two nonzero components: the transverse angular terms $h^{TT}_{\hat{\theta}\hat{\theta}} = -h^{TT}_{\hat{\phi}\hat{\phi}}$ and $h^{TT}_{\hat{\theta}\hat{\phi}} = h^{TT}_{\hat{\phi}\hat{\theta}}$. Using this, we can compute each term in the expression for $\psi_4$ (see Eq. (9.123) in \cite{Baumgarte}). For example, the second term becomes
\begin{equation}
    -2i \, {}^{(4)}R_{\hat{t}\hat{\theta}\hat{t}\hat{\phi}} 
    = i \, \partial_{\hat{t}} \partial_{\hat{t}} h_{\hat{\theta}\hat{\phi}} 
    = i \, \ddot{h}_{\times}, \nonumber
\end{equation}
where the double dots denote the second time derivative. Note that far from the source, in the wave zone, we have the relation $\partial_t h^{\text{TT}}_{bd} = -\partial_r h^{\text{TT}}_{bd}$. Substituting all relevant components into the definition of $\psi_4$, we find
\begin{equation} \label{eq:psi4}
    \psi_4 = \ddot{h}_+ - i \, \ddot{h}_{\times}.
\end{equation}

\vspace{3mm}
It's useful to decompose $\psi_4$ into $-2$ spin-weighted spherical harmonics
\begin{align}
    \psi_4(t, r, \theta, \phi) = 
    \sum_{\ell=2}^\infty \sum_{m=-\ell}^{\ell} 
    \psi_4^{\ell m}(t, r) \; {}_{-2}Y_{\ell m}(\theta, \phi),
\end{align}
where $\psi_4^{\ell m}(t, r)$ are the mode amplitudes, and ${}_{-2}Y_{\ell m}(\theta,\phi)$ are the spin-weighted spherical harmonics with spin weight $s=-2$. The index $\ell$ denotes the angular degree, and $m$ represents the azimuthal mode number \cite{Baumgarte}.

\vspace{3mm}
\section*{Method}
To begin the evolution, we must specify initial data that must satisfy the Hamiltonian and momentum constraint equations (\ref{eq:constraint}). In both of our simulations, we adopt Brill–Lindquist initial data \cite{Brill}, widely used for modeling multiple black holes. This initial data are time-symmetric and assume a conformally flat spatial metric. The conformal factor $\psi$ is given by
\begin{align}
    \psi = 1 + \sum_{i=1}^N \frac{m_{(i)}}{2 \left| \vec{r} - \vec{r}_{(i)} \right|}; \qquad K_{ij} = 0,
\end{align}
where $m_{(i)}$ is the mass parameter of the $i$th black hole, $\vec{r}_{(i)}$ is its coordinate location, and $|\vec{r} - \vec{r}_{(i)}|$ is the Euclidean distance to that black hole.  Additionally to these initital conditions, we choose the initial lapse $\alpha$ and the initial shift $\beta^i$ to be $\psi^{-2}$ and $0$, respectively.

\vspace{3mm}
The evolution equations (\ref{eq:evolution}) and the gauge-condition equations (\ref{eq:gauge}) are first-order in time and can all be expressed in the general form  
\begin{equation}
    \partial_t u(t) = \mathcal{L}(u(t), t),
\end{equation}
where $u = \{\bar{\gamma}_{i j}, \bar{A}_{ij}, \phi, K, \bar{\Lambda}^i, \alpha, \beta^i, B^i\}$ represents the collection of 24 evolved variables. As indicated by equations (\ref{eq:evolution}) and (\ref{eq:gauge}) , the operator $\mathcal{L}$ involves various components of $u$, including their first and second spatial derivatives. Finite difference schemes are used to evaluate the spatial derivatives of the evolved variables on a uniformly sampled grid \cite{IZB}. To integrate this system in time, we use the classical fourth-order Runge–Kutta (RK4) method \cite{RK4}. So, we have
\begin{equation}
    u(t + \Delta t) = u(t) + \frac{1}{6} \left( k_1 + 2k_2 + 2k_3 + k_4 \right),
\end{equation}
where each $k_i$ represents an intermediate stage computed using successive evaluations of the right hand side of the evolution equations. To ensure stability when using RK4, the Courant-Friedrichs-Lewy (CFL) condition must be satisfied. NRPy+ module enforces this by computing the smallest proper distance, $\Delta s_{\min}$, between adjacent grid points across all coordinate directions, (see Eq.(54) in \cite{IZB}), and setting the time step according to $\Delta t = C \Delta s_{\min}$, with the Courant factor  $C = 0.5$ for all simulations presented here.

\vspace{3mm}
The initial data have to satisfy the Hamiltonian and the momentum constraints, meaning that these constraints must be satisfied at the beginning of the time iteration. At each RK4 substep, we evaluate the right hand sides of the evolution equations (\ref{eq:evolution}) and (\ref{eq:gauge}). During each substep, we also apply appropriate boundary conditions.

\vspace{3mm}
In NRPy+, most spatial derivatives are computed using centered finite difference stencils, which require $N_G = \frac{\text{finite difference derivative order}}{2} + 1$ layers of ghost points surrounding the mathematical domain to evaluate derivatives near the boundaries. At the outer boundaries, ghost zone points are filled according to the chosen outer boundary condition, such as the commonly used Sommerfeld condition or quadratic extrapolation. However, approximate boundary conditions can introduce unwanted ingoing modes that affect the simulation interior. Logarithmically spaced radial coordinates allow the outer boundary to be placed sufficiently far away to minimize these effects, keeping it outside causal contact with the origin. Inner boundary conditions depend on the coordinate system and must account for intrinsic periodic, axial, and radial symmetries. In spherical coordinates, the domain is bounded by $0 < r < \infty$, $0 < \theta < \pi$, and $0 < \phi < 2\pi$. In this case, there is only one outer boundary at $r \to \infty$, while the remaining five are inner boundaries. These inner boundaries correspond to radial symmetry about the origin, axial symmetry about the north and south poles, and periodic symmetry around the azimuthal axis in both positive and negative orientations \cite{IZB}.

\vspace{3mm}
Finally, since we require $\partial_t \bar{\gamma} = 0$ \cite{Brown}, where $\bar{\gamma} = \det \bar{\gamma}_{ij}$, the determinant of the conformal metric is expected to remain equal to its initial value. However, numerical errors can cause $\bar{\gamma}$ to drift away from its constant value over time. To correct for this deviation, we adjust $\bar{\gamma}_{ij}$ at the end of each RK4 time step according to 
\begin{equation}
    \bar{\gamma}_{ij} \rightarrow \left( \frac{\hat{\gamma}}{\bar{\gamma}} \right)^{1/3} \bar{\gamma}_{ij} \nonumber
\end{equation}
ensuring that the determinant is restored to $\bar{\gamma} = \hat{\gamma}$, where $\hat{\gamma}$ is the determinant of the reference metric $\hat{\gamma}_{ij}$. 

\vspace{3mm}
After evolving the spacetime variables, we extract the gravitational radiation content by computing the Newman–Penrose scalar $\psi_4$ at fixed coordinate radii. As previously discussed, $\psi_4$ encodes the outgoing gravitational wave signal, see equation (\ref{eq:psi4}), in the wave zone and is evaluated using a null tetrad aligned with spherical coordinates. The extracted waveform is then decomposed into spin-weighted spherical harmonics to analyze individual $(\ell, m)$ modes of the signal.
\newpage
\vspace{3mm}
\section*{Results and Discussions}
In both test cases, we adopted nearly identical numerical setups, differing primarily in two aspects. For the binary black hole merger, both are non-spinning equal-masses $m_1 = m_2 = 0.5M$, while for the three black hole merger, they are non-spinning equal-masses $m_1 = m_2 = m_3 = \frac{1}{3}M$, where $M$ is the total mass. In the binary case, the black holes are aligned axisymmetrically on the $z$-axis, whereas in the three black holes case, the black holes are positioned symmetrically around the origin.

\begin{figure}[ht]
    \centering
    \captionsetup{width=0.8\linewidth}
    \includegraphics[width=70mm]{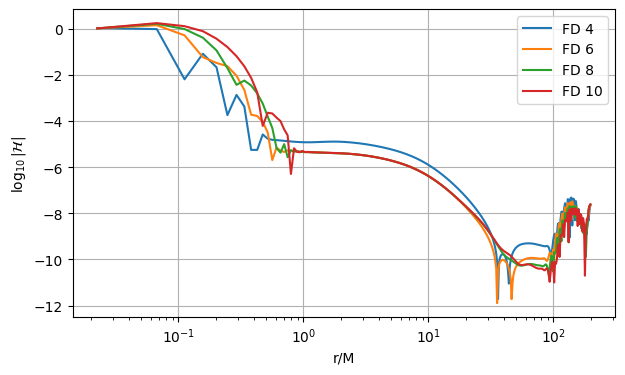}
    \includegraphics[width=70mm]{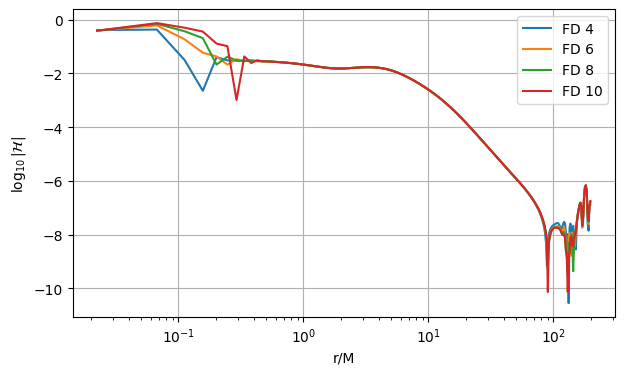} 
    \caption{Logarithmic Hamiltonian constraint violation, $\log_{10}|\mathcal{H}|$, at the final time $t = 200M$ for the binary collision (left plot) and the three-black-holes collision (right plot).}
    \label{fig:Log10HError}
\end{figure}

\vspace{3mm}
For both tests, we used a grid resolution of $[N_r, N_\theta, N_\phi] = [300, 20, 2]$ on spherical coordinate system, where $N_r$ represents the number of grid interior along the radial direction, $N_\theta$ is the number of grid interior along the polar direction, and $N_\phi$ the number of grid interior along the azimuthal direction. With only two points in the azimuthal direction, the 3D simulation effectively reduces to a 2D simulation. The outer boundary is placed at $r_{\text{max}} = 200M$. A width parameter $w = 0.2$ is employed, see Eq. (63) in \cite{IZB}, which controls the clustering of grid points near the origin. Kreiss–Oliger dissipation is applied \cite{IZB}, with dissipation strength $0.3$ for the gauge variables, and $0.3$ for the metric components. The shift damping parameter is set to $\eta = 2/M$. Evolutions are performed on fixed grids using finite difference schemes, denoted by FD $= 4, 6, 8,$ and $10$. The finale time in both test is set to $200M$.

\vspace{3mm}
In Fig. \ref{fig:Log10HError}, the Hamiltonian constraint violation is shown for both the binary and three black hole simulations. As expected, the violation is largest inside the black hole horizons. In the binary case, left plot, the results from $6$th, $8$th, and $10$th order finite difference schemes are nearly indistinguishable. This is, most likely, due to insufficient resolutions for higher order schemes, i.e. $6$th-$10$th orders. In the three black hole case, right plot, all finite difference orders yield virtually identical results. This behavior is again attributed to limited resolution. Due to runtime constraints on Google Colab, we were not able to increase the grid resolution further, otherwise the file would crash when simulations exceed 12 hours of runtime \cite{colab}.

\begin{figure}[ht]
    \centering
    \captionsetup{width=0.8\linewidth}
    \includegraphics[width=70mm]{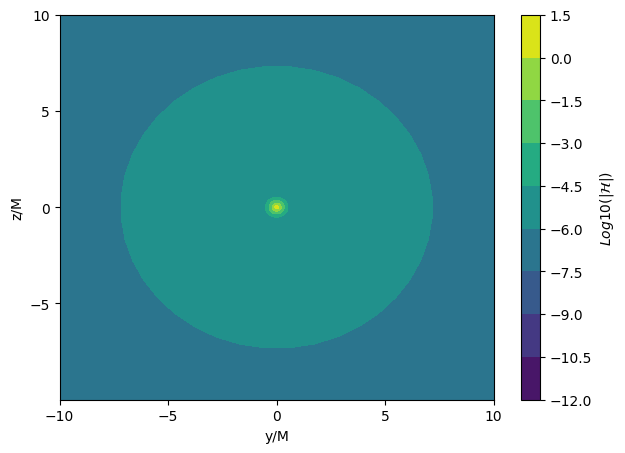}
    \includegraphics[width=70mm]{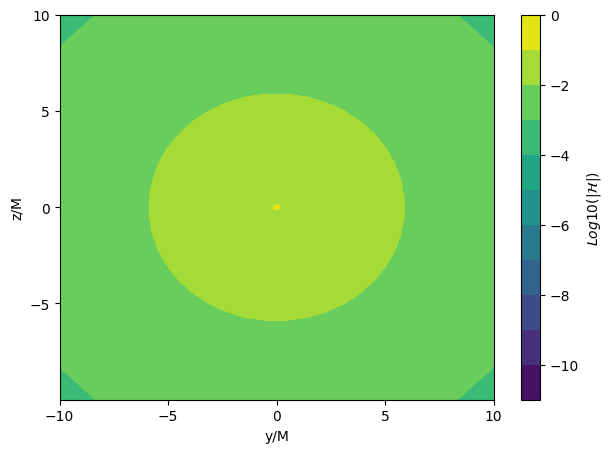} 
    \caption{2D plots for the Hamiltonian constraint violation, $\log_{10}|\mathcal{H}|$, at the final time $t=200M$ for the binary collision (left plot) and the three black holes collision (right plot).}
    \label{fig:2DLog10HError}
\end{figure}

\vspace{3mm}
To illustrate the importance of resolution, we present 2D plots, Fig. \ref{fig:2DLog10HError}, of the Hamiltonian constraint violation with the 6th finite difference order. In both cases, the bright yellow region at the center corresponds to the newly formed black hole. As expected, the violation decreases with distance from the black hole. However, the right plot, corresponding to the three black hole test, exhibits significantly larger violations, indicating that a finer grid resolution is necessary to achieve better results.

\begin{figure}[ht]
    \centering
    \captionsetup{width=0.8\linewidth}
    \includegraphics[width=70mm]{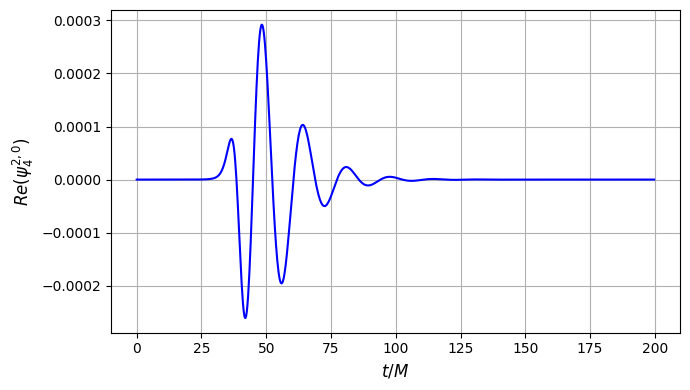} 
    \includegraphics[width=70mm]{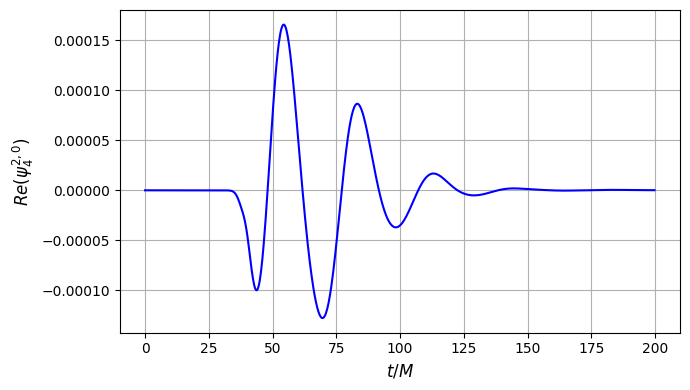} 
    \caption{Dominant waveform mode $(\ell=2, m=0)$ of $Re(\psi_4)$ extracted at $R=30M$ using 6th-order finite differencing. Left: binary black hole merger. Right: trinary black hole merger.}
    \label{fig:waveforms}
\end{figure}

\vspace{3mm}
A collision of black holes results in the emission of gravitational waves. In Fig.~\ref{fig:waveforms}, we plot the dominant mode $(\ell=2, m=0)$ of the real part of $\psi_4$ for both tests. In both cases, the gravitational waveforms were extracted at radius $R_{\text{ext}} = 30M$, using 6th-order finite differencing. We observed that the waveform signal from the binary collision exhibits a larger amplitude and smaller interval, while the signal from the three-black-hole collision shows a lower amplitude and broader interval. The Hamiltonian constraint violation was higher near the black holes in the three-black-hole test, right panel. Thus, the resulting waveform may not be good enough for direct comparison with the binary black hole test, left panel. However, we include it to illustrate how insufficient grid resolution can affect the accuracy of gravitational waveform extraction. It might look acceptable here, but when plotting it alongside other modes, i.e., $\ell=2$ and $m = \pm2, \pm1, 0$, we observed noticeable noise in most modes, see Fig \ref{fig:catalog2} in appendix \ref{appendix:C}.

\begin{figure}[ht]
    \centering
    \captionsetup{width=0.8\linewidth}
    \includegraphics[width=70mm]{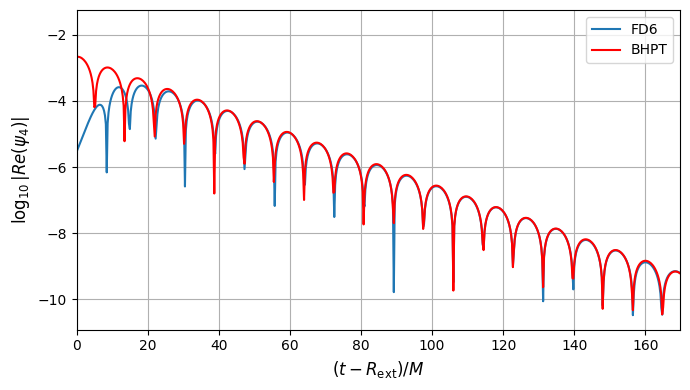} 
    \includegraphics[width=70mm]{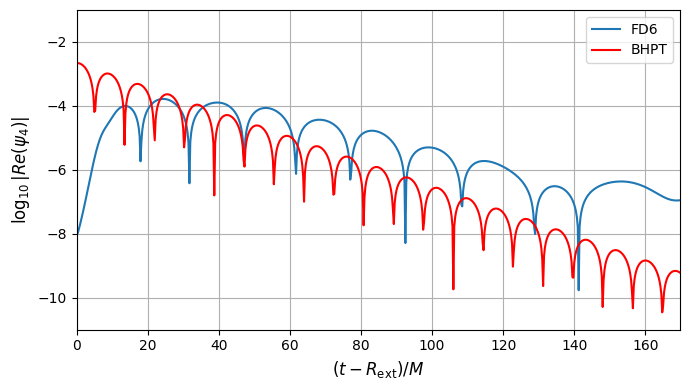} 
    \caption{Log10 of the dominant mode $(\ell = 2, m = 0)$ of $|\text{Re}(\psi_4)|$ using 6th order finite differencing. Left: binary black hole merger matches BH perturbation theory. Right: three black hole merger compared to BH perturbation theory.}
    \label{fig:ringdown}
\end{figure}

To validate our results and assess the physical accuracy of both the evolution and the waveform extraction, we compare the ringdown phase of the dominant mode with analytical predictions from black hole perturbation theory \cite{Berti}. In Fig.~\ref{fig:ringdown}, we illustrate the decay of gravitational waves over time for both tests using 6th order finite differencing. Specifically, we plot the dominant mode $(\ell=2, m=0)$ of the real part of $\psi_4$, extracted at radius $R_{\text{ext}} = 30$, and compare it to the fundamental quasinormal mode (QNM) of a Schwarzschild black hole \cite{Berti}. As shown in Fig.~\ref{fig:ringdown}, the gravitational waveform from the binary black hole test, left plot, exhibits excellent agreement with the expected ringdown signal predicted by black hole perturbation theory. 

\vspace{3mm}
To conclude, the primary objective of this work was to implement and simulate the collision of three black holes using the NRPy+ infrastructure, and to compare the resulting dynamics and gravitational waveforms with those from a binary black hole merger. While the binary test reproduced expected physical behavior and matched predictions from black hole perturbation theory, the three black hole simulation highlighted key challenges. In particular, larger constraint violations and inconsistent waveform extraction revealed the sensitivity of such complex systems to numerical resolution. These results emphasize the necessity of increasing the grid resolution to achieve physically reliable simulations in multi-black-hole scenarios. 

\vspace{3mm}
NRPy+ has demonstrated the ability to simulate both single black holes and binary mergers with Hamiltonian constraint violations that exhibit exponential convergence to zero \cite{IZB}. In our work, the limitations imposed by Google Colab restricted the achievable grid resolution. Nevertheless, for the binary black hole test, we obtained reliable results using 6th order finite differencing, as higher order schemes, 8th and 10th, did not yield any improvements. In contrast, the three black hole collision clearly requires a higher grid resolution to reduce constraint violations and improve the accuracy of the extracted waveforms.

\begin{figure}[ht]
    \centering
    \captionsetup{width=0.8\linewidth}
    \includegraphics[width=150mm]{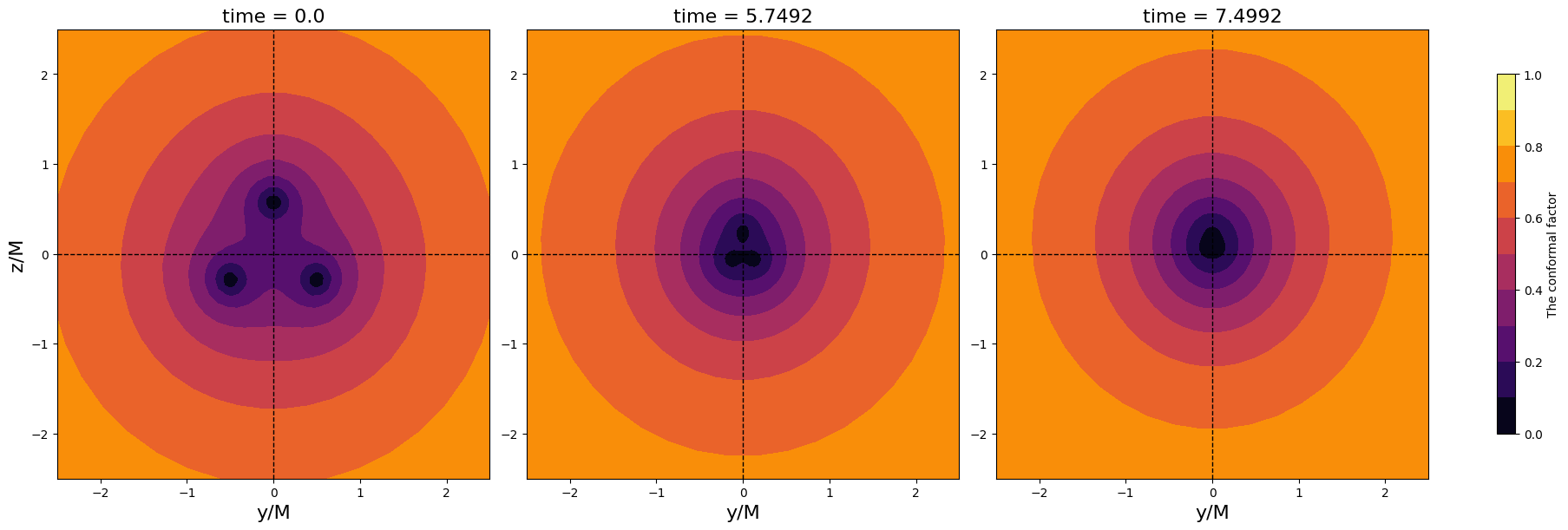} 
    \caption{Simulation of three black holes collision, the initial setup at $t = 0M$, first collide at $t=5.7492M$, and right after the final merger at $t \approx 7.5M$.}
    \label{fig:3bhs}
\end{figure}

\vspace{3mm}
During our attempts to test the implementation of three black holes using NRPy+, we observed that when the black holes were symmetrically arranged around the origin as shown in Fig.~\ref{fig:3bhs}, the two located at the bottom collided first, breaking the symmetry. They subsequently merged with the third black hole positioned above, resulting in a recoil. This is surprising since a symmetric configuration should result in zero net linear momentum; instead, we observed an asymmetric collision producing a nonzero linear momentum \cite{Baumgarte}. This could be explored as a direction for future work, where one may investigate the problem and study impact of the linear momentum on the resulting gravitational waves.

\vspace{3mm}
Now that we understand the limitations of implementing three black hole simulations within Google Colab, and given that our current study focused solely on non-spinning equal-mass black holes, a natural direction for future work would be to study and analyze binary black hole mergers with spinning or non-spinning unequal masses black holes. As a result of such mergers, the remnant black hole may not settle into a static state, but instead experience a recoil, or the kick. Exploring this behavior would open the door to a wide range of future projects in black hole dynamics and gravitational wave extraction.

\vspace{3mm}
\section*{Acknowledgments}
I would like to express my sincere gratitude to my advisor, Dr. Aim\'e Fournier, and my project committee for their guidance and support throughout this work. I am thankful to the university writing center for their help in improving my writing. I also acknowledge the use of AI tools, including AI features in Google Colab for code generation and debugging, and Gemini for assistance in finding sources and clarifying concepts, which supported me in organizing and drafting parts of this project. Finally, I am deeply grateful to my family and friends for their constant encouragement and support.

\newpage
\begin{appendices}
\section{}
\label{appendix:A}
\begin{itemize}
    \item[1.] The line element, which tells us the distance between nearby points, can be written using 3+1 coordinates as:
    \begin{align}
        ds^2 &= g_{ab} dx^a dx^b \\
             &=  -\alpha^2 dt^2 + \gamma_{ij}\left(dx^i + \beta^i dt\right)\left(dx^j + \beta^j dt\right) \, ,
    \end{align}
    where the three-spatial metric $\gamma_{ij}$ is just the projection of the $g_{ab}$ into the hypersurface $\Sigma_t$, Fig. \ref{fig:3+1}, $\beta^i$ shift vector, and $\alpha$ lapse function.
    
    \vspace{3mm}
    \item[2.] The Ricci tensor $R_{ab}$ and Ricci scalar $R$ are derived from the Riemann tensor, see equation (1.19) in \cite{Baumgarte},
    \begin{align}
        R^c_{acb} &= R_{ab} \, ,\\
        R^a_a  &= R \;\; .
    \end{align}
    \vspace{3mm}
    \item[3.] The Christoffel symbols $\bar{\Gamma}^i_{jk}$ 
    \begin{equation}\label{eq:Christoffel}
    \Gamma^{a}{}_{bc} = g^{ad} \, \Gamma_{dbc} 
    = \frac{1}{2} g^{ad} \left( \partial_{c} g_{db} 
    + \partial_{b} g_{dc} 
    - \partial_{d} g_{bc} \right) \; .
    \end{equation}
    \vspace{3mm}
    \item[4.]  The extrinsic curvature $K_{ij}$
    \begin{align} \label{eq:extrinsic}
        K_{ij} = -\frac{1}{2}\mathcal{L}_{\textbf{n}}\gamma_{ij} \,,
    \end{align}
    where $\mathcal{L}_{\textbf{n}}$ is the Lie derivative along the normal vector $n^a$. (See A.1 The Lie derivative in \cite{Baumgarte} and D.1 Differential Geometry in \cite{Baumgarte2} for more about the Lie derivative)
\end{itemize}

\begin{itemize}
    \item[5.] $\partial_{\perp}$is the hypersurface-normal derivative operator, see \cite{BSSNQ},
    \begin{equation}
        \partial_{\perp} = \partial_t - \mathcal{L}_\beta \, ,
    \end{equation}
    
    where $\mathcal{L}_\beta$ is the Lie derivative along the shift vector $\beta^i$.
    \vspace{3mm}
    \item [6.] $\bar{\gamma}_{ij}$ is the conformal metric
    \begin{equation}
        \bar{\gamma}_{ij} = \varepsilon_{i j} + \hat{\gamma}_{ij},
    \end{equation}
    where the reference metric $\hat{\gamma}_{ij}$ is the flat space metric, and $\varepsilon_{i j}$ encodes the deviation of the conformal metric from flat space.
    \vspace{3mm}
    \item[7.] The conformal Ricci tensor $\bar{R}_{ij}$ is computed as:
    \begin{align}
    \bar{R}_{ij} {} = {} & -\frac{1}{2} \bar{\gamma}^{kl} \hat{D}_{k} \hat{D}_{l} \bar{\gamma}_{ij} 
    + \bar{\gamma}_{k(i} \hat{D}_{j)} \bar{\Lambda}^{k} 
    + \Delta^{k} \Delta_{(ij)k} \nonumber \\
    & + \bar{\gamma}^{kl} \left(2 \Delta_{k(i}^{m} \Delta_{j)ml} 
    + \Delta_{ik}^{m} \Delta_{mjl} \right) \; ,
    \end{align}
    
    where:
    \begin{itemize}
        \item $\hat{D}_j$ is the covariant derivative with respect to the reference metric $\hat{\gamma}_{ij}$. 
        \item $\bar{D}_j$ is the covariant derivative with respect to the conformal (barred) spatial 3-metric $\bar{\gamma}_{ij}$. 
        
        (See A.1 The Lie derivative in \cite{Baumgarte} and D.1 Differential Geometry in \cite{Baumgarte2} for more about the covariant derivative)
        
        \item $\Delta^i{}_{jk}$ is the tensor constructed from the difference between the Christoffel symbols of the barred and reference metrics, see equations (6) \& (7) in \cite{IZB}.
        \item The term $\bar{\gamma}_{k(i} \hat{D}_{j)} \bar{\Lambda}^k$ is a shorthand for $
         \frac{1}{2} \left( \bar{\gamma}_{ki} \hat{D}_j \bar{\Lambda}^k + \bar{\gamma}_{kj} \hat{D}_i \bar{\Lambda}^k \right)$.
         \item The term $2 \Delta^m_{k(i} \Delta_{j) m l}$ is a shorthand for $\Delta^m_{k i} \Delta_{j m l} 
        + \Delta^m_{k j} \Delta_{i m l} \; .$
    \end{itemize}
    \vspace{3mm}
    \item[8.]  The term that appears in equation (\ref{eq:evolution}b), $4 \bar{D}_{(i} \alpha \bar{D}_{j)} \phi$, is a shorthand for $2 \bar{D}_i \alpha \, \bar{D}_j \phi + 2 \bar{D}_j \alpha \, \bar{D}_i \phi$.
    \vspace{3mm}
    \item[9.] The advective time derivative $\partial_0 = \partial_t - \beta^i \partial_i$.
\end{itemize}

\newpage
\section{}
\label{appendix:B}

-   All terms in this appendix are expressed in terms of the metric perturbation

\vspace{3mm}
1. The inverse metric is given by:
\begin{equation}
    g^{ab} = \eta^{ab} + k^{ab}
\end{equation}

Or, in terms of $ h^{ab}$:
\begin{align}
g_{ac}g^{cb} &= (\eta_{ac} + h_{ac})(\eta^{cb} + k^{cb}) \nonumber\\
\delta^{b}_{a} &= \delta^{b}_{a} + \eta_{ac}k^{cb} + h_{ac}\eta^{cb} \\
\Rightarrow \qquad k^{rb} &= - h^{br} \qquad \text{(after multiplying by } \eta^{ra}) \nonumber
\end{align}
\begin{equation}
    \boxed{g^{ab} = \eta^{ab} - h^{ab}}
\end{equation}

2. The Christoffel symbols
\begin{align}
\Gamma^{c}_{ab} &= \frac{1}{2} g^{cd}
\left( \partial_{b} h_{da} + \partial_{a} h_{db} - \partial_{d} h_{ab} \right) \nonumber \\
&= \frac{1}{2} \left( \eta^{cd} - h^{cd} \right)
\left( \partial_{b} h_{da} + \partial_{a} h_{db} - \partial_{d} h_{ab} \right) \\
&= \boxed{\frac{1}{2} \eta^{cd}
\left( \partial_{b} h_{da} + \partial_{a} h_{db} - \partial_{d} h_{ab} \right)} \nonumber
\end{align}

3. Riemann Tensor
\begin{align}
R^{d}_{cab} &= \partial_{a} \Gamma^{d}_{bc} - \partial_{b} \Gamma^{d}_{ac} 
+ \Gamma^{e}_{bc} \Gamma^{d}_{ae} - \Gamma^{f}_{ac} \Gamma^{d}_{bf} \nonumber\\
&= \frac{1}{2} \eta^{dg} 
\left( \partial_{a} \partial_{c} h_{gb} 
- \partial_{a} \partial_{g} h_{bc} 
- \partial_{b} \partial_{c} h_{ga} 
+ \partial_{b} \partial_{g} h_{ac} \right) \\
\eta_{ed} R^{d}_{cab} &= 
\frac{1}{2} \eta_{ed} \eta^{dg} 
\left( \partial_{a} \partial_{c} h_{gb} 
- \partial_{a} \partial_{g} h_{bc} 
- \partial_{b} \partial_{c} h_{ga} 
+ \partial_{b} \partial_{g} h_{ac} \right) \nonumber
\end{align}

\[
\boxed{
\begin{aligned}
R_{ecab} = \frac{1}{2} \Big(
&\partial_{a} \partial_{c} h_{eb} +
\partial_{b} \partial_{e} h_{ac} -
\partial_{b} \partial_{c} h_{ea} -
\partial_{a} \partial_{e} h_{bc} \Big)
\end{aligned}
}
\]

4. The Einstein tensor in linearized gravity:
\begin{equation}
    G_{ab} = R_{ab} - \frac{1}{2} \eta_{ab} R
\end{equation}

With the Ricci tensor:
\begin{align}
    R_{ab} = \frac{1}{2} \left( 
\partial_{a} \partial_{c} h^{c}{}_{b} + 
\partial_{b} \partial_{d} h^{d}{}_{a} - 
\Box h_{ab} - 
\partial_{a} \partial_{b} h 
\right)
\end{align}

Ricci scalar:
\begin{equation}
    R = \partial_{a} \partial_{b} h^{ab} - \Box h
\end{equation}

Wave operator (d'Alembertian)
\begin{equation} \label{eq:d'A}
    \Box = \eta^{ab} \partial_{a} \partial_{b}
\end{equation}

Trace of the perturbation
\begin{equation}
    h = \eta^{ab} h_{ab}
\end{equation}

Trace-reversed perturbation
\begin{equation} \label{eq:tracerp}
    \boxed{\bar{h}_{ab} = h_{ab} - \frac{1}{2} \eta_{ab} h}
\end{equation}

Then the linearized Einstein equations become
\[
\boxed{
\begin{aligned}
G_{ab} = \frac{1}{2} (&
\partial^{c} \partial_{a} \bar{h}_{cb} + 
\partial^{c} \partial_{b} \bar{h}_{ac} - 
\Box \bar{h}_{ab} - 
\eta_{ab} \partial^{c} \partial^{d} \bar{h}_{cd})
\end{aligned}
}
\]

5. Derivative of the Plane Wave
\begin{align}
    \partial_{c} (\bar{h}_{ab})
    &= \partial_{c} \left( A_{ab} e^{i k_{d} x^{d}} \right) \nonumber \\
    &= A_{ab} \, \partial_{c} \left( e^{i k_{d} x^{d}} \right) \nonumber\\
    &= A_{ab} \, e^{i k_{d} x^{d}} \, \partial_{c} (k_{d} x^{d}) \nonumber\\
    &= A_{ab} \, e^{i k_{d} x^{d}} \, i k_{d}\delta^{d}_{c} \nonumber\\
    &= A_{ab} \, e^{i k_{d} x^{d}} \, i k_{c} \nonumber\\
    &= \boxed{\bar{h}_{ab} \, i k_{c}}
\end{align}

6. The Lorenz Gauge Implies
\begin{align}
0 &= \partial_{a}\bar{h}^{ab} \nonumber\\
  &= \partial_{a}\left(A^{ab} e^{i k_{c} x^{c}}\right) \nonumber \\
  &= \boxed{i\,A^{ab}\,k_{a}\,e^{i k_{c} x^{c}}}
\end{align}

\newpage
\section{}
\label{appendix:C}
\begin{figure}[ht]
    \centering
    \includegraphics[width=160mm]{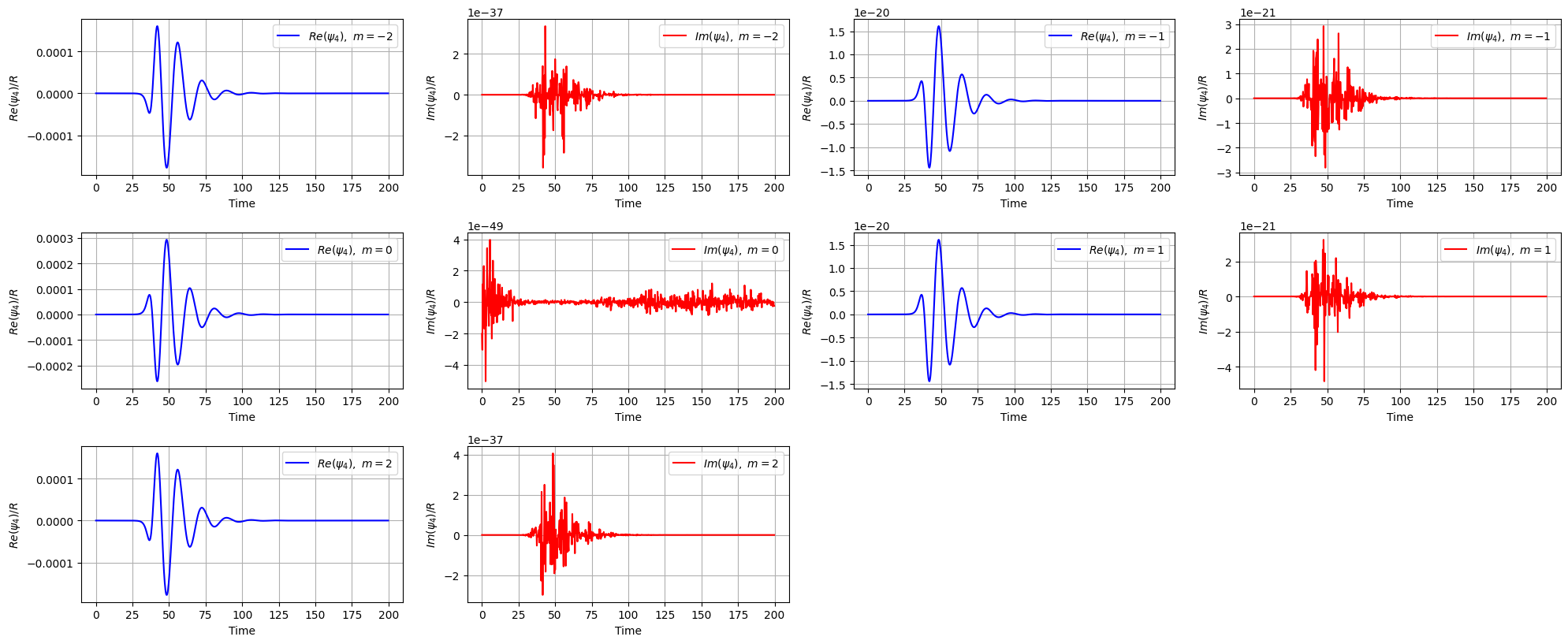}
    \caption{Gravitational wave extraction showing the modes $(\ell = 2, m = 0, \pm1, \pm2)$ from the binary black hole simulation. The real part of $\psi_4$ is shown in blue and the imaginary part in red. The imaginary components for $m = \pm2, 0$, second column, are negligible compared to their corresponding real components ,first column. The $m = \pm1$ modes (third and fourth columns) are significantly smaller than the  real modes $m = \pm2, 0$, but have comparable amplitudes to each other.}
    \label{fig:catalog1}
\end{figure}
\newpage
\begin{figure}[ht]
    \centering
    \includegraphics[width=160mm]{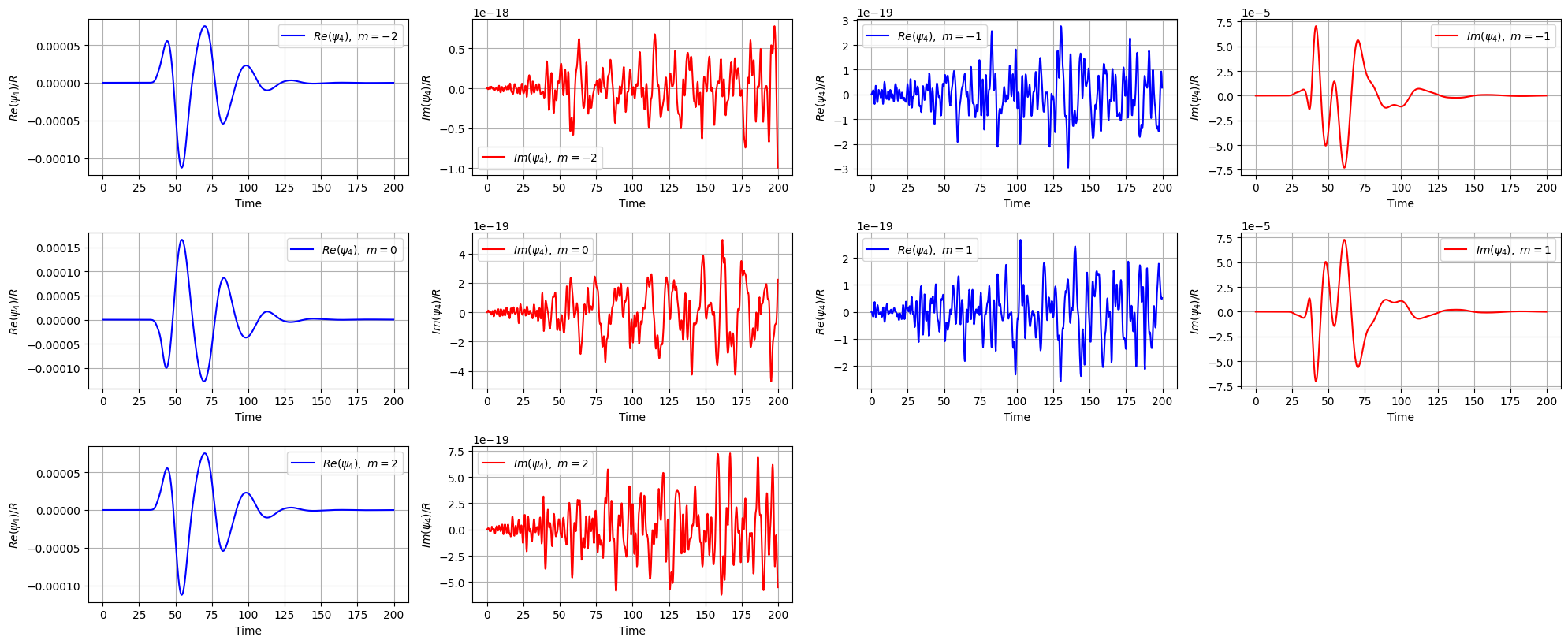} 
    \caption{Gravitational wave extraction showing the modes $(\ell = 2, m = 0, \pm1, \pm2)$ from the three-black-hole simulation. The real part of $\psi_4$ is shown in blue and the imaginary part in red. The imaginary components for $m = \pm2, 0$, second column, are negligible compared to their corresponding real parts, first column. The real components for the $m = \pm1$ modes, third column, are negligible, have significantly lower amplitude, compared to their corresponding imaginary components, fourth column. }
    \label{fig:catalog2}
\end{figure}
\end{appendices}
\newpage
\printbibliography

@book{Baumgarte,
  author    = {Thomas W. Baumgarte and Stuart L. Shapiro},
  title     = {Numerical Relativity: Solving Einstein's  Equations on the Computer},
  year      = {2010},
  publisher = {Cambridge University Press},
  address   = {Cambridge}
}

@book{Baumgarte2,
  author    = {Thomas W. Baumgarte and Stuart L. Shapiro},
  title     = {Numerical Relativity Starting from Scratch},
  year      = {2021},
  publisher = {Cambridge University Press},
  address   = {Cambridge}
}

@book{carroll,
  author    = {Sean M. Carroll},
  title     = {Spacetime and Geometry: An Introduction to General Relativity},
  year      = {2014},
  publisher = {Cambridge University Press},
  address   = {United Kingdom}
}

@misc{etienneBHaH,
  author = {Zachariah B. Etienne},
  title  = {{BlackHoles@Home}},
  url    = {https://blackholesathome.net/nrpy.html},
  note   = {}
}

@article{Brill,
    author = {Brandt, S. R. and Bruegmann, B.},
    title = {{BH Punctures as Initial Data for General Relativity}},
    eprint = {gr-qc/9711015},
    archivePrefix = {arXiv},
    primaryClass = {gr-qc},
    doi = {},
    year = {1997}
    }

@misc{RK4,
  author       = {{Wikipedia contributors}},
  title        = {Runge--Kutta methods --- Wikipedia{,} The Free Encyclopedia},
  year         = {2019},
  url          = {https://en.wikipedia.org/w/index.php?title=Runge%E2%80%93Kutta_methods&oldid=898536315},
  note         = {}
}

@misc{BSSNQ,
  author       = {Zachariah Etienne},
  title        = {nrpytutorial: Tutorial-BSSN\_quantities.ipynb},
  year         = {},
  howpublished = {\url{https://github.com/zachetienne/nrpytutorial/blob/master/Tutorial-BSSN_formulation.ipynb}},
  note         = {}
}

@article{IZB,
  author    = {Ian Ruchlin and Zachariah B. Etienne and Thomas W. Baumgarte},
  title     = {SENR/NRPy+: Numerical relativity in singular curvilinear coordinate systems},
  journal   = {},
  year      = {2018},
  url       = {https://journals-aps-org.aurarialibrary.idm.oclc.org/prd/abstract/10.1103/PhysRevD.97.064036}
}

@article{Nerozzi,
  author       = {Andrea Nerozzi and Christopher Beetle and Marco Bruni and Lior M. Burko and Denis Pollney},
  title        = {Towards wave extraction in numerical relativity: the quasi-Kinnersley frame},
  journal      = {Physical Review D},
  volume       = {72},
  year         = {2005},
  doi          = {10.1103/PhysRevD.72.024014},
  url = {https://journals-aps-org.aurarialibrary.idm.oclc.org/prd/abstract/10.1103/PhysRevD.72.024014}
}

@article{Brown,
  author       = {J. David Brown},
  title        = {Covariant Formulations of {BSSN} and the Standard Gauge},
  journal      = {Physical Review D},
  year         = {2009},
  doi          = {},
  eprint       = {},
  doi       = {10.1103/PhysRevD.79.104029},  
  url       = {https://journals-aps-org.aurarialibrary.idm.oclc.org/prd/pdf/10.1103/PhysRevD.79.104029}
}

@misc{colab,
  author       = {{Google Research}},
  title        = {{Google Colaboratory FAQ}},
  howpublished = {\url{https://research.google.com/colaboratory/faq.html}},
  note         = {},
  year         = {2025}
}

@article{Berti,
  author    = {Emanuele Berti and Vitor Cardoso and Andrei O. Starinets},
  title     = {Quasinormal modes of black holes and black branes},
  journal   = {Classical and Quantum Gravity},
  year      = {2009},
  doi       = {10.1088/0264-9381/26/16/163001},
  url       = {https://iopscience-iop-org.aurarialibrary.idm.oclc.org/article/10.1088/0264-9381/26/16/163001}
}

@article{Palenzuela,
  author    = {Carlos Palenzuela},
  title     = {Introduction to Numerical Relativity},
  journal   = {},
  year      = {2020},
  url       = {https://arxiv.org/abs/2008.12931}
}

@article{Schnetter,
  author       = {Erik Schnetter},
  title        = {Time Step Size Limitation Introduced by the BSSN Gamma Driver},
  journal      = {Classical and Quantum Gravity},
  year         = {2010},
  archivePrefix = {arXiv},
  primaryClass = {gr-qc},
  url          = {https://iopscience-iop-org.aurarialibrary.idm.oclc.org/article/10.1088/0264-9381/27/16/167001}
}
\end{document}